\documentclass[twocolumn,times,trackchanges, dvipsnames]{aastex631}
\usepackage[utf8]{inputenc}
\usepackage{amsmath}

\begin{document}
\title{Robust support for semi-automated reductions of Keck/NIRSPEC data using PypeIt}

\author{Adolfo S. Carvalho}
\affiliation{Department of Astronomy; California Institute of Technology; Pasadena, CA 91125, USA}
\author{Greg Doppmann}
\affiliation{W. M. Keck Observatory, 65-1120 Mamalahoa Hwy, Kamuela, HI, USA}
\author[0000-0003-1809-6920]{Kyle B.~Westfall}
\affiliation{University of California Observatories, University of
  California, Santa Cruz, 1156 High St., Santa Cruz, CA 95064, USA}
\author{Debora Pelliccia}
\affiliation{University of California Observatories, University of
  California, Santa Cruz, 1156 High St., Santa Cruz, CA 95064, USA}
\author{J. Xavier Prochaska}
\affiliation{University of California, Santa Cruz; 1156 High St., Santa Cruz, CA 95064, USA}
\author{Joseph Hennawi}
\affiliation{Physics Department, Broida Hall, University of California, Santa Barbara, CA 93106-9530, USA}
\affiliation{Leiden Observatory, Leiden University, PO Box 9513, 2300 RA Leiden, The Netherlands}
\author{Frederick B. Davies}
\affiliation{Max-Planck-Institut für Astronomie, Königstuhl 17, D-69117 Heidelberg, Germany}
\author[0009-0008-9808-0411]{Max Brodheim}
\affiliation{W. M. Keck Observatory, Waimea, HI 96743, USA}
\author{Feige Wang}
\affiliation{Department of Astronomy, University of Michigan, 1085 S. University Ave., Ann Arbor, MI 48109, USA}
\affiliation{Steward Observatory, University of Arizona, 933 N Cherry Avenue, Tucson, AZ 85721, USA}
\author[0000-0001-7653-5827]{Ryan Cooke}
\affiliation{Centre for Extragalactic Astronomy, Durham University, Durham, DH1 3LE, UK}

\begin{abstract}
    We present a data reduction pipeline (DRP) for Keck/NIRSPEC built as an addition to the PypeIt Python package. The DRP is capable of reducing multi-order echelle data taken both before and after the detector upgrade in 2018. As part of developing the pipeline, we implemented major improvements to the capabilities of the PypeIt package, including manual wavelength calibration for multi-order data and new output product that returns a coadded spectrum order-by-order. We also provide a procedure for correcting telluric absorption in NIRSPEC data by using the spectra of telluric standard stars taken near the time of the science spectra. At high resolutions, this is often more accurate than modeling-based approaches. 
\end{abstract}

\section{Introduction}
\label{sec:introduction}

The high resolution Near InfraRed SPECtrograph \citep[NIRSPEC,][]{McLean_nirspecDesign_1998SPIE} on the W.~M.~Keck Observatory delivers a resolving power $R \equiv \lambda/\Delta \lambda = 15,000-35,000$ from 0.95 to 5.5 $\mu$m. Prior to 2018, there were three recommended software options for reducing NIRSPEC data: the IDL package REDSPEC\footnote{\url{https://www2.keck.hawaii.edu/inst/nirspec/redspec.html}}, the IRAF package WMKONSPEC\footnote{\url{https://www2.keck.hawaii.edu/inst/nirspec/wmkonspec.html}}, and the semi-automated Python pipeline NSDRP\footnote{\url{https://www2.keck.hawaii.edu/koa/nsdrp/nsdrp.html}}. In 2018, the instrument underwent a major detector upgrade, in which the 1024x1024 detector was replaced by a 2048x2048 detector. Post-upgrade, only REDSPEC was updated to reduce the data from the new detector, rendering NSDRP and WMKONSPEC usable only for pre-upgrade data. 

In order to enable NIRSPEC users to rapidly reduce new and archival data, we used the existing data reduction pipeline package PypeIt \citep{prochaska_pypeit_2020JOSS}. 
PypeIt is a python-based data-reduction package that enables automated processing of raw on-sky observations and calibration data from slit-based spectrographs.  Separate workflows are implemented for long-slit, multi-slit, slicer-based integral-field units, and cross-dispersed echelle observations; however, all workflows leverage the same core code to perform common tasks.  Instrument-specific code is isolated from the core code in a way that limits the development needed to introduce new spectrographs into the PypeIt architecture.  Although \textit{enabling} PypeIt to read and apply its routines to data from additional spectrographs is streamlined, ensuring that the PypeIt reductions are \textit{reliable and robust} requires dedicated development effort.

In this Note, we detail this effort for PypeIt reductions of Keck/NIRSPEC data, applicable to data obtained both before and after the detector upgrade in 2018. The pipeline includes improvements to the main PypeIt code and a procedure for the removal of telluric absorption in NIRSPEC spectra using telluric standard star spectra. The pipeline is available at the PypeIt Github page (\url{https://github.com/pypeit/PypeIt}) and an example script outlining the telluric removal procedure is provided in the Keck/NIRSPEC instrument webpage and linked in Section \ref{sec:telluric}.

\section{Updates to PypeIt}

Developing the PypeIt-NIRSPEC module required two major upgrades to the PypeIt main data reduction code. The first, and most impactful, was the expansion of the PypeIt manual wavelength calibration program to operate on echelle data. The second was the construction of a new data structure as a default output for PypeIt reductions of echelle data. Here, we will briefly describe the two upgrades. 

The original user-interface-based manual wavelength calibration in PypeIt was written for calibrating only a single trace, which is appropriate in the case of long-slit spectroscopy or for multi-object spectroscopy where each trace is expected to have approximately the same wavelength solution. To wavelength calibrate the multi-trace echelle spectra, the script now iterates through each extracted order, prompts the user to compute a solution for the order through a user-interface, then saves the solution and proceeds to the next order. Upon completing all of the orders, the multi-trace solution is saved so that it may be used for any data set taken with the same grating settings. The script was tested with data and calibrations taken at different times with independent instrument configurations. It can handle spectral shifts and spatial shifts (8 and 25 pix, respectively) typical of the variation in the positioning accuracy of NIRSPEC's gratings.

The second major update to PypeIt was in the output products. Previously, the final co-added output spectrum, even for high resolution echelle spectrographs, was presented in a single masked vector spanning the entire wavelength range of the spectrum. While this is ideal for low and medium resolution spectroscopy, particularly when the spectra are properly flux-calibrated, this can make working with high resolution spectra challenging. Now, for high resolution echelle spectra, a second output product is offered: the co-added spectrum given as an $N \times M$ array for the $N$ orders of the spectrum and $M$ columns on the detector. This is an output format that is standard in other data reduction pipelines for high resolution spectra, including NSDRP. 

\section{Telluric Contamination Correction}\label{sec:telluric}

The recommended telluric correction technique for NIRSPEC data is to use standard A0V star spectra taken throughout the night at varying airmasses to ensure accurate removal of atmospheric absorption features. To then remove the telluric features from the observed spectrum requires the following steps:
\begin{enumerate}
    \item Continuum normalizing the standard star and science target spectra.
    \item Fitting and dividing out the stellar photospheric features of the (ideally A0V) standard.
    \item Shifting the newly-obtained telluric absorption spectrum to match the science target in case of instrumental wavelength shifts between observations.
    \item Uniform scaling the line depths of the telluric absorption spectrum to account for differences in airmass between observations.
    \item Dividing the shifted and scaled telluric absorption spectrum from the science target spectrum. 
\end{enumerate}

\begin{figure*}
    \centering
    \includegraphics[width = 0.45\linewidth]{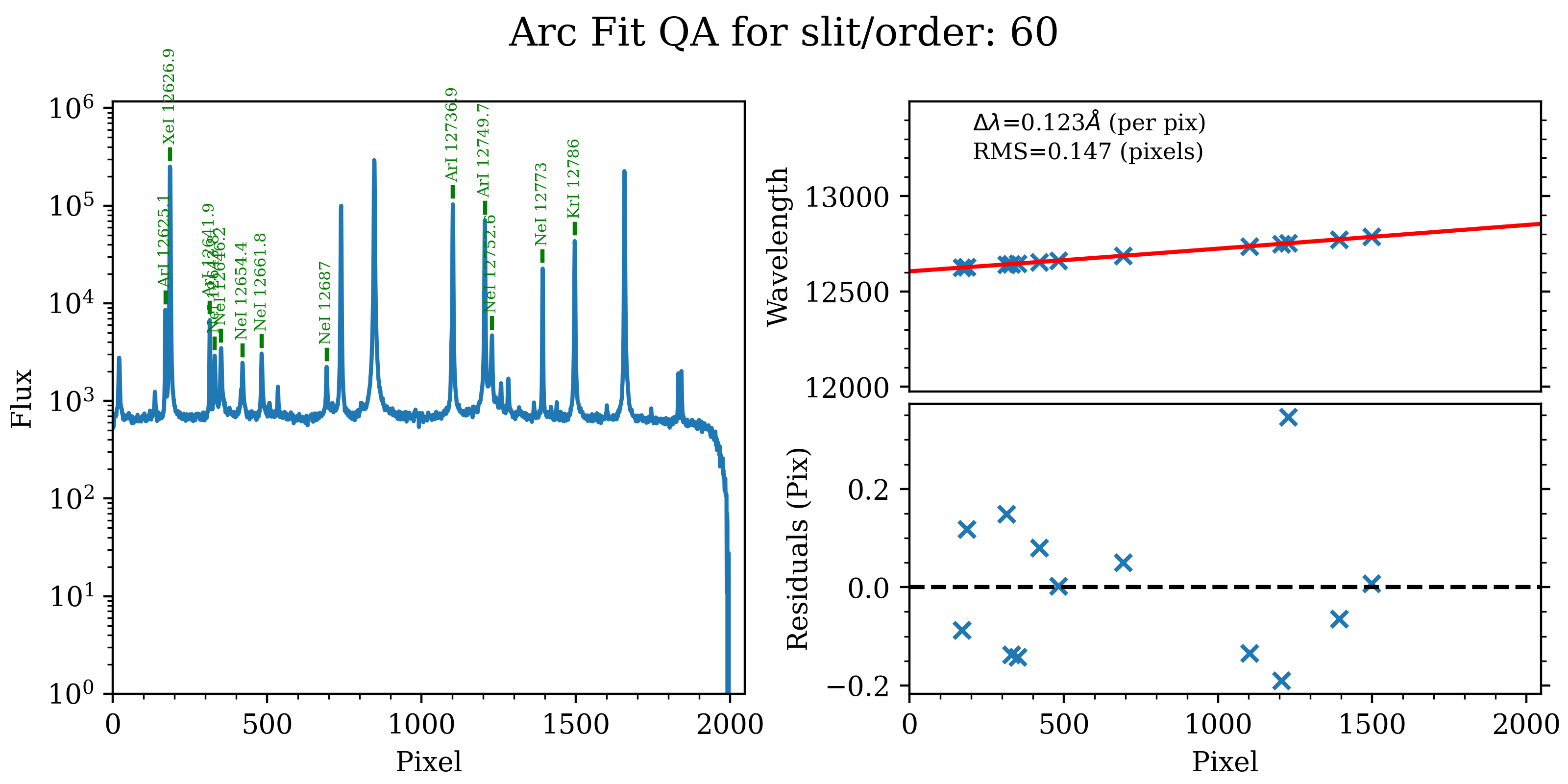}
    \includegraphics[width = 0.45\linewidth]{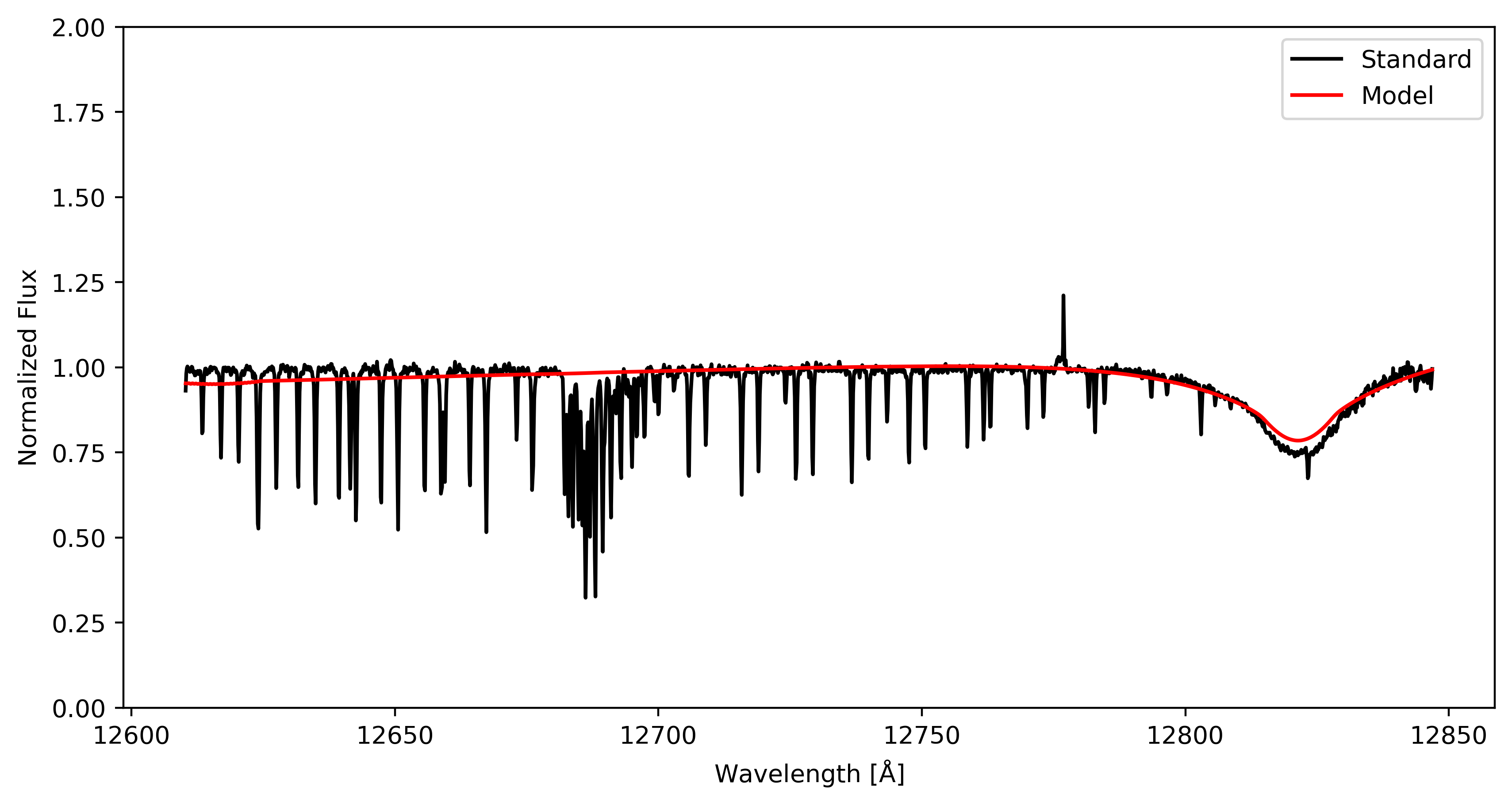}
    \includegraphics[width=0.98\linewidth]{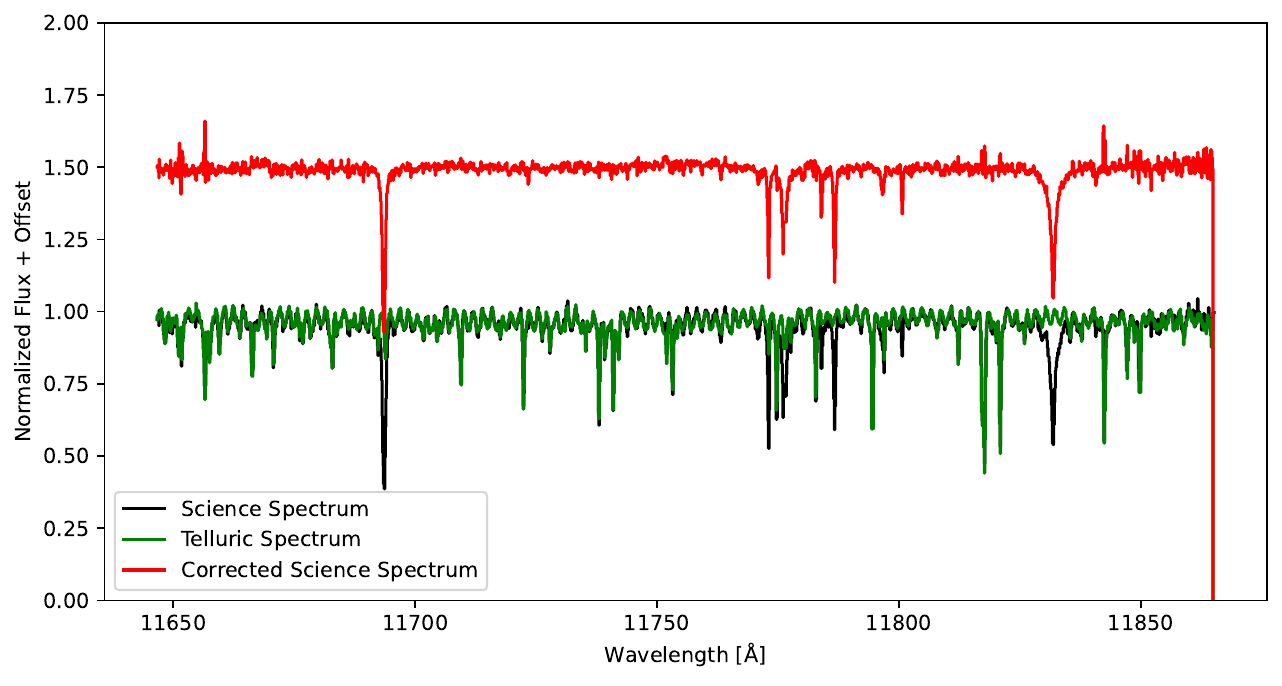}
    \caption{\textbf{Upper Left:} A sample quality assurance plot from the wavelength calibration of Order 60 in the J band. \textbf{Upper Right:} Order 60 of the A0V telluric standard spectrum (black) and the PHOENIX model (red) used to remove the Pa$\beta$ line to produce a telluric absorption spectrum. \textbf{Bottom Panel:} A plot of Order 65 of the science spectrum (black), the telluric spectrum (green) and the telluric-corrected spectrum (red). Notice that the fringing is also removed by the division by the telluric absorption spectrum. }
    \label{fig:TellCorr}
\end{figure*}

We make use of several tools in the PypeIt package for this procedure, particularly the cross-correlation-based approach used to compute the relative shifts between spectra. A well-documented notebook showing how to extract the output PypeIt data and implement the steps above is provided at \url{https://www2.keck.hawaii.edu/inst/nirspec/pypeit_nirspec.html} to guide the user. We recommend this as a means of telluric correction, which is based on the popularly-used telluric correction program in IRAF. An example showcasing the efficacy of our telluric correction is shown in Figure \ref{fig:TellCorr}. 

In the Y and J bands, observations taken using the thin blocker filter are also subject to severe fringing that is worse in bluer orders. The fringing is stable as a function of wavelength over the course of the night and appears in both the standard and science spectra at the same amplitude relative to continuum. As such, using a standard star for telluric correction in the same night as the observed science target enables easy removal of the fringing as well. This is demonstrated in the case of the J band spectrum presented in Figure \ref{fig:TellCorr}.

If standard star observations are not available for a particular night or a desired spectrograph setup, PypeIt offers a modeling-based means of telluric line removal. However, the model-based removal is less reliable at the highest resolutions and performs poorly in Y and J band spectra that have significant fringing.

\section{Acknowledgements}
\software{Astropy \citep{astropy_2013, astropy_2018, astropy_2022}, NumPy \citep{harris2020array}}

\bibliography{references}{}
\bibliographystyle{aasjournal}



\end{document}